\documentclass[12pt,prd,tightenlines,nofootinbib,showpacs,showkeys]{revtex4}
\usepackage{bm}
\usepackage{color,graphics}
\usepackage{rotating}
\usepackage{epsfig}
\usepackage{amssymb}
\usepackage{amsmath}
\usepackage{amsthm}
\usepackage{amsfonts}
\usepackage{amssymb}
\usepackage{braket}
\usepackage{indentfirst}
\usepackage{epstopdf}
\usepackage[colorlinks=true,linkcolor=blue,citecolor=blue,urlcolor=blue]{hyperref}

\begin{document}

\title{Energy levels of mesonic helium in quantum electrodynamics}
\author{\firstname{V.~I.}~\surname{Korobov}}
\affiliation{BLTP JINR, Dubna, Russia}
\affiliation{Samara National Research University, Samara, Russia}
\author{\firstname{A.~V.}~\surname{Eskin}}
\affiliation{Samara National Research University, Samara, Russia}
\author{\firstname{A.~P.}~\surname{Martynenko}}
\affiliation{Samara National Research University, Samara, Russia}
\author{\firstname{F.~A.}~\surname{Martynenko}}
\affiliation{Samara National Research University, Samara, Russia}

\begin{abstract}
On the basis of variational method we study energy levels of pionic helium $(\pi-e-He)$ and kaonic helium 
$(K-e-He)$ with an electron in ground
state and a meson in excited state with principal and orbital quantum numbers $n\sim\l+1\sim 20$.
Variational wave functions are taken in the Gaussian form.
Matrix elements of the basic Hamiltonian and corrections to vacuum polarization and
relativism are calculated analytically in a closed form.
We calculate some bound state energies and transition frequencies
which can be studied in the experiment.
\end{abstract}



\maketitle

\section{Introduction}
\label{vv}

One of the directions in the development of the theory of fundamental interactions is connected with a study 
of bound states of particles. In addition to usual stable atoms and molecules that exist in our world, 
there are exotic bound states (muonium, positronium, positronium ion, muonic hydrogen, and others), which have 
attracted the attention of both experimenters and theoreticians for decades \cite{egs,aldo,famu}. 
Although they have a short lifetime, 
nevertheless, by studying various energy intervals in the energy spectrum of such systems, as well as their decay 
widths, year after year it was possible to obtain from these studies more accurate information about the values 
of fundamental parameters of the Standard Model. A number of such exotic systems has been growing in recent 
years. For example, in \cite{hori1,hori2}, it was proposed to study by laser spectroscopy method
pionic helium atoms, which consist of a negative pion, 
an electron, and a helium nucleus. From a measurement of pion transitions between 
states with large values of the principal and orbital quantum numbers $((n,l)=(17.16)\to (17.15))$ one can try 
to obtain a more accurate value of the pion mass than can be done by other methods. In \cite{hori3,hori4}, a successful 
experiment has already been carried out for nearly circular orbits $n\sim l+1$, which gave a transition frequency 
value of 183760 MHz. To find a more 
accurate value of the pion mass from these measurements, it is also necessary to take into account
systematic effects such as collision induced shift, broadening of the transition lines and others 
\cite{bakalov1,bulg,bakalov}.
The work in this direction is in an active phase. Along with the atoms of pionic helium, other 
atoms can be proposed and studied, for example, kaonic helium, setting as the goal of research a more accurate 
determination of the mass of the $K^{-}$ meson.
It will be useful to note that there are other approaches to clarifying a value of the $\pi$ meson mass. 
Thus, the study carried out in \cite{crystal} demonstrates the potential of crystal spectroscopy of curved crystals 
in the field of exotic atoms.
In this work, $5g-4f$ transitions in pionic nitrogen and muonic oxygen were measured simultaneously in 
a gaseous nitrogen-oxygen mixture. Knowing the muon mass, the muon line can be used to energy calibrate 
the pion transition. The mass value of negatively charged pion was obtained, which is 4.2 ppm higher 
than the current world average $139.57077\pm 0.00017$ MeV \cite{pdg}.

Mesonic atoms are formed as a result of a replacement of an orbital electron by a negatively charged meson. 
After that, laser spectroscopy of such atoms is carried out, which will make it possible 
to measure transition frequencies and determine the reduced mass of a system and hence a mass of the meson.
To reduce the influence of strong interaction between a meson and a nucleus, the meson's orbit 
is raised by increasing its orbital momentum.
The long lifetime of a meson atom is determined by the state with a large value of orbital momentum $l=(16\div 20)$
in which a meson is formed in the atom.
Its transition to the ground state with $l=0$ is strongly suppressed. 
The lifetime of such an atom is several nanoseconds.

The study of energy levels of three-particle systems can be carried out with high accuracy 
within the framework of the variational method.
There are some differences in the use of a variational method to find the energy levels of three-particle 
systems. They are connected with a choice of coordinates and representation of the Hamiltonian to describe 
the system, with a choice of basis wave functions.  Thus, in \cite{hori1} an exponential basis was used, and the coordinates 
of the electron and meson are determined with respect to the nucleus. In works 
\cite{phys2023,bul2023,apm2023}, 
when calculating the energy levels of mesomolecules of hydrogen, muonic helium, etc.,
we use the Jacobi coordinates. The purpose of this work is to calculate the energy levels in pionic 
and kaonic helium atoms, as well as transition frequencies between levels in which the meson is in 
an excited state with a large orbital quantum number.

\section{General formalism}
\label{gf}

Different approaches have been developed for a study of three-particle systems. 
There is an analytical method of perturbation theory, which makes it possible to analytically 
investigate both the Lamb shift and the hyperfine structure of the spectrum 
\cite{mohr,huang,amusia,amusia1,apm2008,apm2022,apm2023}. Another methods that are used 
for many-particle systems are the variational method and method of hyperspherical coordinates, 
which allow one to find energy levels and wave functions 
with very high accuracy \cite{rd,melezhik,frolov,frolov1,chen,iran,varga,korobov,khan}. Since for mesonic helium the states 
of an atom with large values of orbital moments 
of the meson are considered so that the electron and meson are at the same distance from the nucleus, 
it is virtually impossible to use a method of analytical perturbation theory. Therefore, further we study 
this system on the basis of the variational method. The Gaussian basis is used as the basis set 
of wave functions.

To find the energy levels of a three-particle system, we introduce the Jacobi coordinates $\boldsymbol\rho$, 
$\boldsymbol\lambda$, which are related to the particle radius vectors ${\bf r}_1$ (nucleus), 
${\bf r}_2$ (meson), ${\bf r}_3$ (electron) as follows:
\begin{equation}
{\boldsymbol\rho}={\bf r}_2-{\bf r}_1,~~~{\boldsymbol\lambda}={\bf r}_3-\frac{m_1{\bf r}_1+m_2{\bf r}_2}{m_1+m_2},
\label{eq1}
\end{equation}
where $m_1$, $m_2$, $m_3$ are the masses of $He$ nucleus, $\pi^-$ ($K^-$)-meson and electron.

To solve the variational problem, we choose the ground state trial basis wave functions in the form 
of superposition of the Gaussian exponents:
\begin{equation}
\Psi({\boldsymbol\rho},{\boldsymbol\lambda},A)=\sum_{i=1}^K C_i \psi_i({\boldsymbol\rho},{\boldsymbol\lambda},A^i),
~~~\psi_i({\boldsymbol\rho},{\boldsymbol\lambda},A^i)=e^{-\frac{1}{2}\left(A_{11}^i{\boldsymbol\rho}^2+
2A_{12}^i{\boldsymbol\rho}{\boldsymbol\lambda}+A_{22}^i{\boldsymbol\lambda}^2\right)},
\label{eq2}
\end{equation}
where $C_i$ are linear variational parameters, $A^i$ is the matrix of nonlinear variational parameters,
K is the basis size. 

In nonrelativistic approximation the Hamiltonian of a three-particle atom in the Jacobi coordinates 
can be presented as
\begin{equation}
\hat H_0=-\frac{1}{2\mu_1}\nabla^2_{\boldsymbol\rho}-\frac{1}{2\mu_2}\nabla^2_{\boldsymbol\lambda}+
\frac{e_1e_2}{|{\boldsymbol\rho}|}+\frac{e_1e_3}{|{\boldsymbol\lambda}+\frac{m_2}{m_{12}}{\boldsymbol\rho}|}+
\frac{e_2e_3}{|{\boldsymbol\lambda}-\frac{m_1}{m_{12}}{\boldsymbol\rho}|},
\label{eq3}
\end{equation}
where $m_{12}=m_1+m_2$, $\mu_1=\frac{m_1m_2}{m_1+m_2}$, $\mu_2=\frac{(m_1+m_2)m_3}{m_1+m_2+m_3}$, $e_1$, $e_2$, $e_3$ are the particle charges. 

For arbitrary states of the meson and electron with orbital angular momenta $l_1$ and $l_2$, a convenient 
basis for the expansion of functions depending on two directions are bipolar sphertic harmonics \cite{var}:
\begin{equation}
[Y_{l_1}(\theta_\rho,\phi_\rho)\otimes Y_{l_2}(\theta_\lambda,\phi_\lambda)]_{LM}=
\sum_{m_1,m_2}C^{LM}_{l_1m_1l_2m_2}Y_{l_1m_1}(\theta_\rho,\phi_\rho)Y_{l_2m_2}(\theta_\lambda,\phi_\lambda),
\label{eq4a}
\end{equation}
where $\theta_\rho,\phi_\rho$ and $\theta_\lambda,\phi_\lambda$ are spherical angles that determine 
the direction of the vectors $\boldsymbol\rho$, $\boldsymbol\lambda$.
Since the $\pi^-$ or $K^-$ meson are in an orbital excited state $l$ in 
pionic (kaonic) helium, and the electron is in the ground state
the variational wave function of the system is chosen for such states in the form:
\begin{equation}
\Psi_{lm}({\boldsymbol\rho},{\boldsymbol\lambda},A)=\sum_{i=1}^K C_i 
Y_{lm}(\theta_\rho,\phi_\rho)\rho^l
e^{-\frac{1}{2}\left(A_{11}^i{\boldsymbol\rho}^2+
2A_{12}^i{\boldsymbol\rho}{\boldsymbol\lambda}+A_{22}^i{\boldsymbol\lambda}^2\right)},
\label{eq4}
\end{equation}
where spherical function $Y_{lm}(\theta_\rho,\phi_\rho)$ describes 
the angular part of the orbital motion of a pion (kaon).

Within the framework of the variational approach, the solution of the Schr\"odinger equation is reduced 
to solving the following matrix problem for the coefficients $C_i$:
\begin{equation}
H\cdot C=EB\cdot C,
\label{eq5}
\end{equation}
where the matrix elements of the Hamiltonian $H_{ij}$ and normalizations $B_{ij}$ can be calculated analytically 
in a basis of the Gaussian wave functions. Thus, the normalization of the wave function \eqref{eq4} is determined 
by the following expression:
\begin{equation}
<\Psi|\Psi>=\sum_{i,j=1}^K C_i C_j 2^{l+2}\pi^{3/2}\Gamma\left(l+\frac{3}{2}\right)
\frac{B_{22}^l}{(det B)^{l+\frac{3}{2}}},~~~
B_{kn}=A_{kn}^i+A_{kn}^j,
\label{eq6}
\end{equation}
where $\Gamma(l+3/2)$ is the Euler gamma function.

Consider further analytical results for the matrix elements of the Hamiltonian.
The kinetic energy operator contains two terms. The matrix element from the Laplace operator with respect 
to ${\boldsymbol\lambda}$ has the form:
\begin{equation}
<\Psi|\nabla^2_{\boldsymbol\lambda}|\Psi>=\sum_{i,j=1}^K C_i C_j 2^{l+2}\pi^{\frac{3}{2}}\Gamma\left(l+\frac{3}{2}\right)
\frac{B_{22}^{l-1}}{(det B)^{l+\frac{5}{2}}}\times
\label{eq7}
\end{equation}
\begin{displaymath}
\left[
3A_{22}^i(A_{22}^i-B_{22})det B+(2l+3)(A_{22}^iB_{12}-A_{12}^iB_{22})^2\right].
\end{displaymath}

Similar matrix element with the Laplace operator in ${\boldsymbol\rho}$ 
is also expressed in terms of nonlinear variational parameters as follows:
\begin{equation}
<\Psi|\nabla^2_{\boldsymbol\rho}|\Psi>=\sum_{i,j=1}^K C_i C_j 2^{l+1}\pi^{\frac{3}{2}}\Gamma\left(l+\frac{1}{2}\right)
\frac{B_{22}^{l-1}}{(det B)^{l+\frac{5}{2}}}\times
\label{eq8}
\end{equation}
\begin{displaymath}
\left[
(2l+1)det B(-(2l+3)A_{11}^iB_{22}+3(A_{12}^i)^2+2l A_{12}^iB_{12})+(2l+1)(2l+3)(A_{12}^iB_{12}-A_{11}^iB_{22})^2
\right].
\end{displaymath}

The potential energy operator in nonrelativistic Hamiltonian consists of pairwise Coulomb interactions
$U_{ij}$ (i, j=1, 2, 3).
The convenience of using the Gaussian basis in this case also lies 
in the possibility of analytical representation of the matrix elements of potential energy
(in electronic atomic units):
\begin{equation}
<\Psi|U_{12}|\Psi>=-Z\sum_{i,j=1}^K C_i C_j 2^{l+\frac{3}{2}}\pi^{\frac{3}{2}}\Gamma\left(l+1\right)
\frac{B_{22}^{l-1}}{(det B)^{l+1}},
\label{eq9}
\end{equation}
\begin{equation}
<\Psi|U_{13}|\Psi>=-Z\sum_{i,j=1}^K C_i C_j 2^{l+\frac{5}{2}}\pi\Gamma\left(l+\frac{3}{2}\right)
\frac{B_{22}^{l+\frac{1}{2}}}{(det B)^{l+\frac{3}{2}}}
{_2F_1}\left(\frac{1}{2},l+\frac{3}{2},\frac{3}{2},-\frac{(F_2^{23})^2}{det B}\right)
\label{eq10}
\end{equation}
\begin{equation}
<\Psi|U_{23}|\Psi>=\sum_{i,j=1}^K C_i C_j 2^{l+\frac{5}{2}}\pi\Gamma\left(l+\frac{3}{2}\right)
\frac{B_{22}^{l+\frac{1}{2}}}{(det B)^{l+\frac{3}{2}}}
{_2F_1}\left(\frac{1}{2},l+\frac{3}{2},\frac{3}{2},-\frac{(F_2^{13})^2}{det B}\right)
\label{eq11}
\end{equation}
\begin{equation}
F_2^{13}=B_{12}+\frac{m_1}{m_{12}}B_{22},~~~F_2^{23}=B_{12}-\frac{m_2}{m_{12}}B_{22},
\label{eq12}
\end{equation}
where ${_2F_1}(\alpha,\beta,x)$ is a hypergeometric function.

For $l=1$ the expressions \eqref{eq10}-\eqref{eq12} coincide with previously obtained results \cite{apm2019}.
Using the matrix elements of the $\hat H_0$ hamiltonian, some energy levels of the $\pi^-$-meson and $K^-$-meson 
atoms are calculated in Matlab system. 
The calculations are carried out using our program, which was previously used to calculate the energy levels 
of various muonic atoms in quantum electrodynamics.
The calculation of the energy levels of the $\pi^-$-meson atom is carried out in order to test the operation of the program.
The calculation results are shown in Table~\ref{tb1}.

To improve the accuracy of the calculation, we consider some important corrections to the Hamiltonian $\hat H_0$.
The pair electromagnetic interaction between particles in quantum electrodynamics is determined by 
the Breit potential \cite{t4}. Among the various terms in this potential, let us single out those terms that 
have the greatest numerical value. These include relativistic corrections, contact interaction and 
corrections for vacuum polarization.

The relativistic corrections are defined in the energy spectrum by the following terms in electronic atomic units:
\begin{equation}
\Delta U_{rel}=-\frac{\alpha^2}{8}\left(\frac{{\bf p}_1^4}{m_1^3}+\frac{{\bf p}_2^4}{m_2^3}+
\frac{{\bf p}_3^4}{m_3^3}\right).
\label{eq13}
\end{equation}

The term of leading order in \eqref{eq13} is related with a motion of the electron.
The value of the matrix element from $\Delta U^e_{rel}$ can be obtained in exactly the same way as \eqref{eq7}
in terms of variational parameters:
\begin{equation}
<\Psi|-\frac{\alpha^2}{8}\nabla^4_{\boldsymbol\lambda}|\Psi>=-\frac{\alpha^2}{8}\sum_{i,j=1}^K C_i C_j 
2^{l+1}\pi^{\frac{1}{2}}\Gamma\left(l+\frac{3}{2}\right)
\frac{B_{22}^{l-2}}{(det B)^{l+\frac{7}{2}}}\Bigl[15(A_{22}^i)^2 (det B)^2(A_{22}^i-B_{22})^2+
\label{eq14}
\end{equation}
\begin{displaymath}
10 (2l+3)  A_{22}^i(A_{22}^i-B_{22})det B(A_{22}^iB_{12}-A_{12}^i B_{22})^2+
(2l+3)(2l+5)(A_{22}^iB_{12}-A_{12}^iB_{22})^4\Bigr].
\end{displaymath}

\begin{table}[htbp]
\caption{\label{tb1} Energy levels of the meson atom obtained in nonrelativistic approximation 
with the Gaussian and exponential basis
and values of main corrections in the energy spectrum in electron atomic units (e.a.u.).
}
\bigskip
\begin{tabular}{|c|c|c|c|c|c|}   \hline
State& $E_{nr}$(Exp) & $E_{nr}$(G)    & $-\frac{\alpha^2}{8}{\bf p}_e^4$  & 
$\Delta U_{vp}$ & $\Delta U_{cont}$ \\
   &  &  &    &     &          \\             \hline
   \multicolumn{6}{|c|}{$({_2^3}He-\pi^-- e)$ atom}     \\            \hline
  (17,16)&-2.64312261030188(2)\cite{hori1}  & -2.6423822152   & -0.0000568853 & -0.0000003596   & 0.0000021185   \\     \hline
 (17,15)&-2.6709980910(1)\cite{hori1} & -2.6698284795 & -0.0000578381 & -0.0000003646 & 0.0000021477     \\    \hline
   \multicolumn{6}{|c|}{$({_2^4}He-\pi^-- e)$ atom}     \\            \hline    
 (17,16)&-2.65751243850171\cite{hori1}   &-2.6567689659   & -0.0000560957    & -0.0000003549   &  0.0000020904  \\     \hline
 (17,15)& -2.68542722(2)\cite{hori1}  &-2.6842422023   & -0.0000571739     & -0.0000003606  & 0.0000021242  \\    \hline
   \multicolumn{6}{|c|}{$({_2^3}He- K^-- e)$ atom}     \\            \hline
 (20,19)& -4.6685977528  &-4.6806222136    & -0.0000194273  &-0.0000001272    & 0.0000007495 \\            \hline
 (20,18)& -4.6786693864  &-4.6932218265    & -0.0000178752  &-0.0000001159    & 0.0000006829 \\               \hline
 (21,20)& -4.3199030879  &-4.3133610685    & -0.0000238657  &-0.0000001466    & 0.0000008638  \\     \hline
 (21,19)& -4.3285605454  &-4.3238629798    & -0.0000230792  &-0.0000001498    & 0.0000008827     \\           \hline    
   \multicolumn{6}{|c|}{$({_2^4}He- K^-- e)$ atom}     \\            \hline
 (20,19)& -4.8266672526  &-4.8328936568   & -0.0000190789   & -0.0000001231    &  0.0000007252      \\      \hline
 (20,18)& -4.8520548999  &-4.8578478878   & -0.0000172623   & -0.0000001161    &  0.0000006552     \\      \hline
 (21,20)& -4.4573174498  &-4.4499843007   & -0.0000207687   & -0.0000001490    &  0.0000008777       \\     \hline
 (21,19)& -4.4653630953  &-4.4601685694   & -0.0000219706   & -0.0000001409    &  0.0000008299    \\    \hline    
\end{tabular}
\end{table}

Let us also take into account the vacuum polarization effects in the energy spectrum.
Since for both an electron and a meson in a highly excited state, the Compton wavelength of an electron 
is much smaller than the radius of the Bohr orbit, we can use the following expression for the vacuum 
polarization potential in electronic atomic units:
\begin{equation}
\Delta U_{vp}=\Delta U_{vp}(r_{13})+\Delta U_{vp}(r_{23})=
-\frac{4}{15}\alpha^2(Z\alpha)\delta({\boldsymbol\lambda}+\frac{m_2}{m_{12}}{\boldsymbol\rho})
+\frac{4}{15}\alpha^2(Z\alpha)\delta({\boldsymbol\lambda}-\frac{m_1}{m_{12}}{\boldsymbol\rho}).
\label{eq15}
\end{equation}

The matrix elements of such potentials are calculated analytically in a closed form:
\begin{equation}
<\Psi|\Delta U_{vp}(r_{13})|\Psi>=-\frac{4}{15}\alpha^2(Z\alpha\sum_{i,j=1}^K C_i C_j 2^{l+\frac{1}{2}}\Gamma\left(l+
\frac{3}{2}\right)\frac{1}{\left(F_1^{13}\right)^{l+\frac{3}{2}}},
\label{eq16}
\end{equation}
\begin{equation}
<\Psi|\Delta U_{vp}(r_{23})|\Psi>=-\frac{4}{15}\alpha^2(Z\alpha\sum_{i,j=1}^K C_i C_j 2^{l+\frac{1}{2}}\Gamma\left(l+
\frac{3}{2}\right)\frac{1}{\left(F_1^{23}\right)^{l+\frac{3}{2}}},
\label{eq17}
\end{equation}
\begin{equation}
F_1^{13}=B_{11}+\frac{m_2^2}{m_{12}^2}B_{22}-2\frac{m_2}{m_{12}}B_{12},~~~
F_1^{23}=B_{11}+\frac{m_1^2}{m_{12}^2}B_{22}+2\frac{m_1}{m_{12}}B_{12}.
\label{eq18}
\end{equation}

The contact interaction potential, as well as \eqref{eq15}, is expressed through the $\delta$-functions 
in the form (in electronic atomic units):
\begin{equation}
\Delta U_{cont}=
\frac{\pi Z\alpha^2}{2}\delta({\boldsymbol\lambda}+\frac{m_2}{m_{12}}{\boldsymbol\rho})
-\frac{\pi \alpha^2}{2}\delta({\boldsymbol\lambda}-\frac{m_1}{m_{12}}{\boldsymbol\rho}).
\label{eq18a}
\end{equation}

In Table~\ref{tb1} we present the results of calculating the energy values with the 
Hamiltonian $\hat H_0$ and the values of the matrix elements \eqref{eq14}, \eqref{eq16}, \eqref{eq17}. \eqref{eq18a}.

\begin{figure}[htbp]
\centering
\includegraphics[scale=0.6]{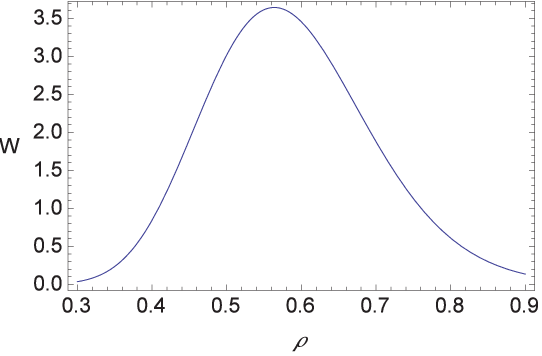}
\includegraphics[scale=0.6]{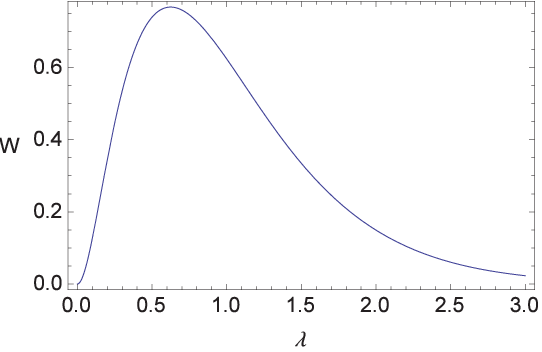}
\caption{The radial distribution densities $W(\rho)$, $W(\lambda)$ for $({^3_2He}-\pi^--e)$
for the state $(16,15)$.
The variable values $\rho$ and $\lambda$ are taken in electron atomic units.}
\label{pic1}
\end{figure}
\begin{figure}[htbp]
\centering
\includegraphics[scale=0.6]{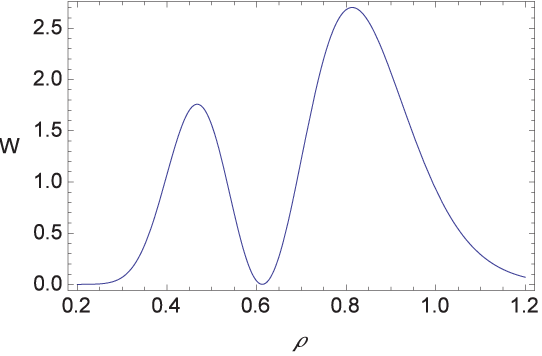}
\includegraphics[scale=0.6]{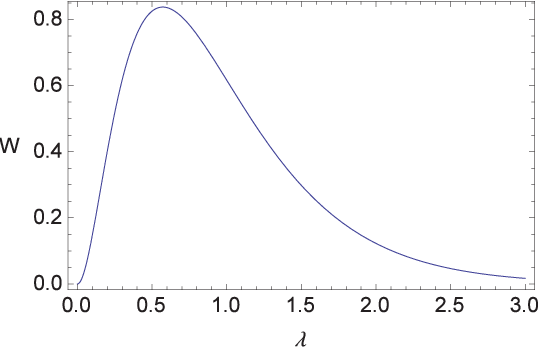}
\caption{The radial distribution densities $W(\rho)$, $W(\lambda)$ for $({^3_2He}-\pi^--e)$
for the state $(17,15)$.
The variable values $\rho$ and $\lambda$ are taken in electron atomic units.}
\label{pic2}
\end{figure}
\begin{figure}[htbp]
\centering
\includegraphics[scale=0.6]{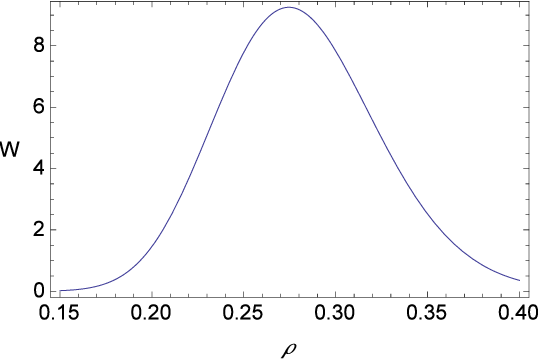}
\includegraphics[scale=0.6]{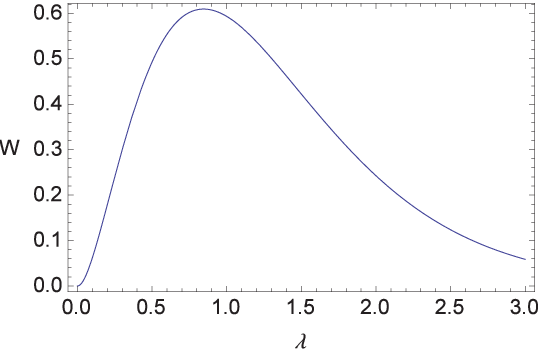}
\caption{The radial distribution densities $W(\rho)$, $W(\lambda)$ for $({^3_2He}-K^--e)$
for the state $(21,20)$.
The variable values $\rho$ and $\lambda$ are taken in electron atomic units.}
\label{pic3}
\end{figure}
\begin{figure}[htbp]
\centering
\includegraphics[scale=0.6]{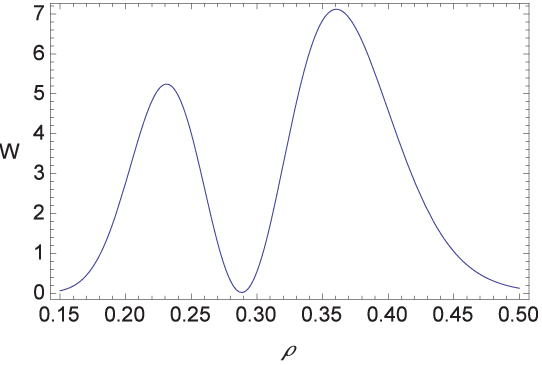}
\includegraphics[scale=0.6]{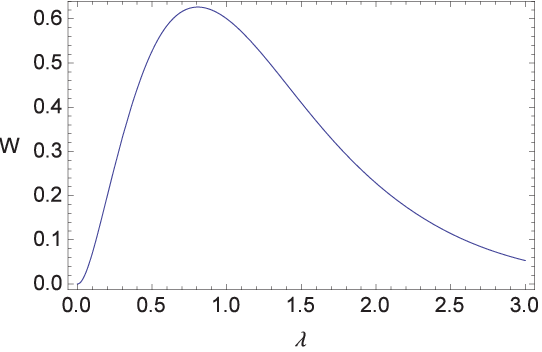}
\caption{The radial distribution densities $W(\rho)$, $W(\lambda)$ for $({^3_2He}-K^--e)$
for the state $(22,20)$.
The variable values $\rho$ and $\lambda$ are taken in electron atomic units.}
\label{pic4}
\end{figure}

The obtained wave functions \eqref{eq4} make it possible to calculate 
the radial distribution densities in $\rho$ and $\lambda$ and  root mean square values
$\sqrt{<\rho^2>}$, $\sqrt{<\lambda^2>}$, which are determined by the expressions:
\begin{equation}
\label{eq19}
W(\rho)=\frac{(2\pi)^{3/2}}{<\Psi|\Psi>}\sum_{i,j=1}^K \frac{C_i C_j}{B_{22}^{3/2}}\rho^{(2l+2)}
e^{-\frac{1}{2}\frac{det B}{B_{22}}\rho^2},
\end{equation}
\begin{equation}
\label{eq20}
W(\lambda)=\frac{2^{l+\frac{5}{3}}\pi}{<\Psi|\Psi>}\sum_{i,j=1}^K \frac{C_i C_j\Gamma\left(l+\frac{3}{2}\right)}
{B_{11}^{l+\frac{3}{2}}}\lambda^2
e^{-\frac{1}{2}B_{22}\lambda^2}{_1F_1}\left(l+\frac{3}{2},\frac{3}{2},\frac{B_{12}^2\lambda^2}{2B_{11}}\right),
\end{equation}
\begin{equation}
\label{eq21}
W(\rho,\lambda)=\frac{4\pi}{<\Psi|\Psi>}\sum_{i,j=1}^K \frac{C_i C_j}{B_{12}}\rho^{2l+1}\lambda
e^{-\frac{1}{2}[B_{11}\rho^2+B_{22}\lambda^2]}sh(B_{12}\rho\lambda),~ B_{lk}=A^i_{lk}+A^j_{lk},
\end{equation}
\begin{equation}
\label{eq22}
<\rho^2>=
\frac{\pi^{\frac{3}{2}}2^{l+3}\Gamma\left(l+\frac{5}{2}\right)}{<\Psi|\Psi>}\sum_{i,j=1}^K C_i C_j\frac{B_{22}^{l+1}}
{(det B)^{l+5/2}},
\end{equation}
\begin{equation}
\label{eq23}
<\lambda^2>=
\frac{\pi^{\frac{3}{2}}2^{l+2}\Gamma\left(l+\frac{3}{2}\right)}{<\Psi|\Psi>}\sum_{i,j=1}^K C_i C_j
\frac{B_{22}^{l-1}}{(det B)^{l+\frac{5}{2}}} (3B_{11}B_{22}+2B_{12}^2l).
\end{equation}

\begin{figure}[htbp]
\centering
\includegraphics[scale=0.35]{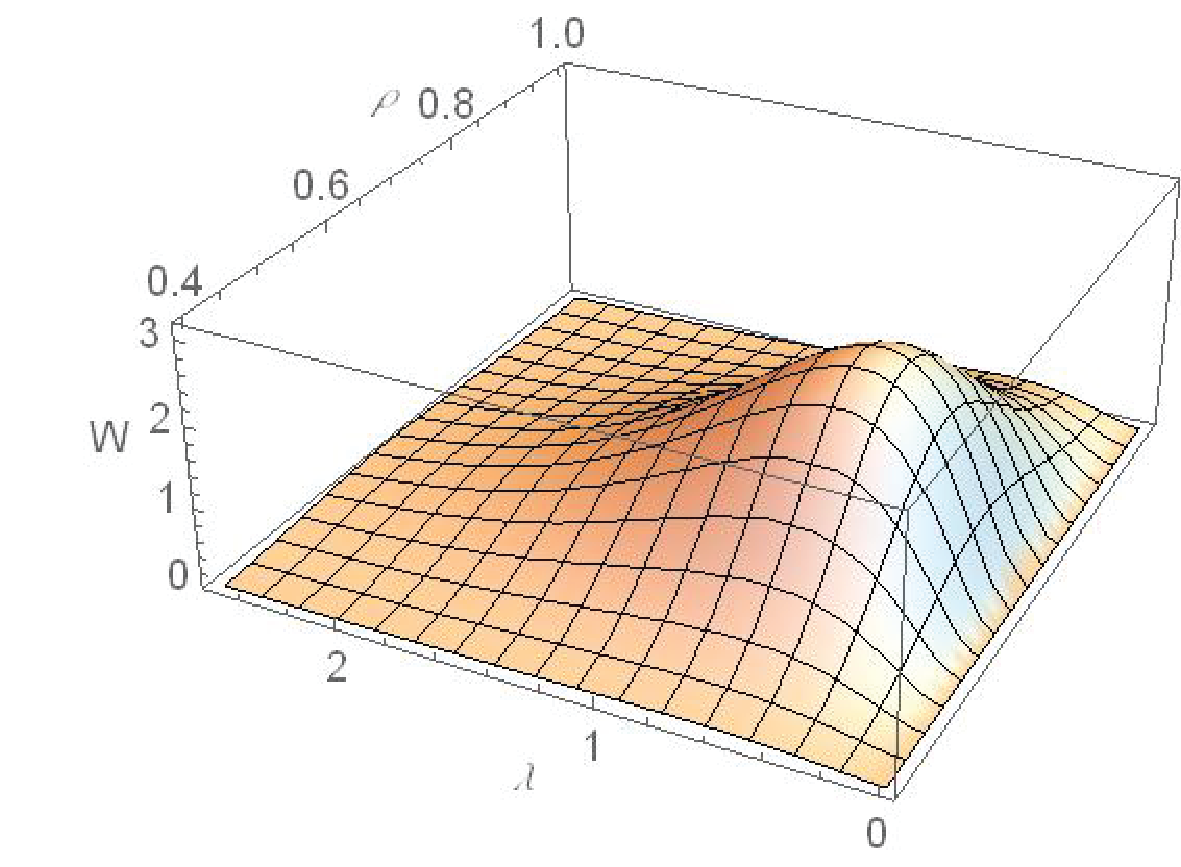}
\includegraphics[scale=0.35]{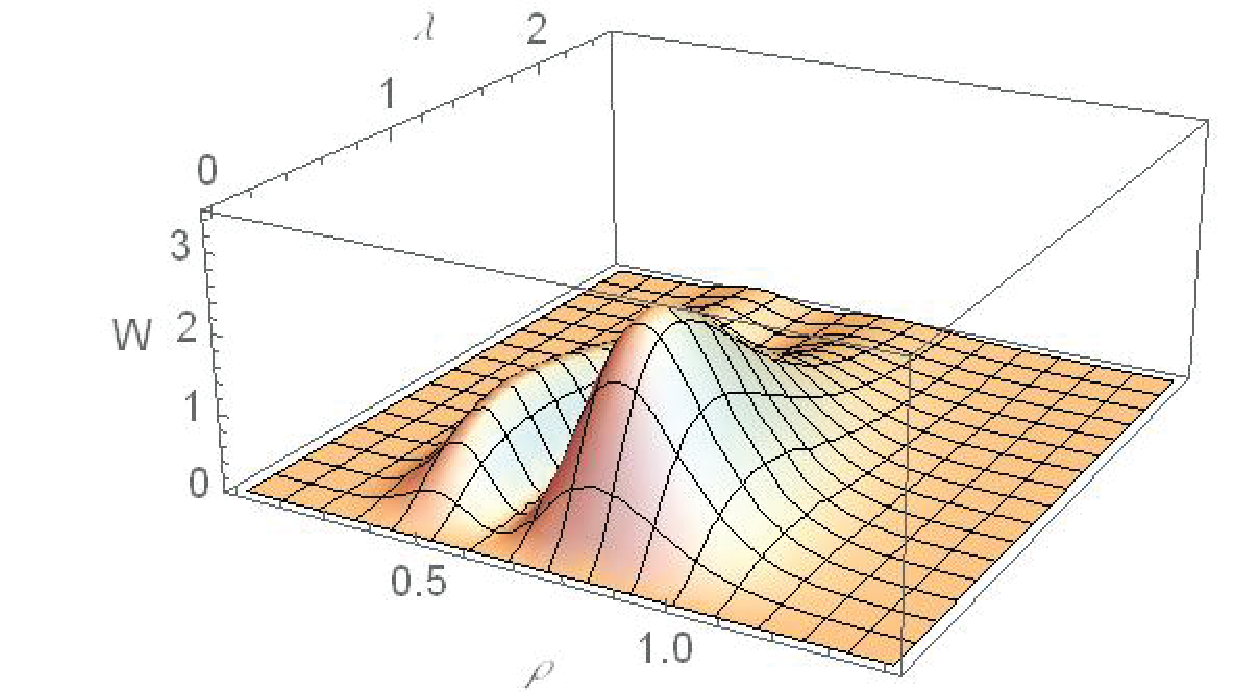}
\caption{The radial distribution density $W(\rho,\lambda)$ for $(\pi e ^4He)$
in states (17,16) and (17,15).
The variable values $\rho$ and $\lambda$ are taken in electron atomic units.
}
\label{pic5}
\end{figure}

\begin{figure}[htbp]
\centering
\includegraphics[scale=0.4]{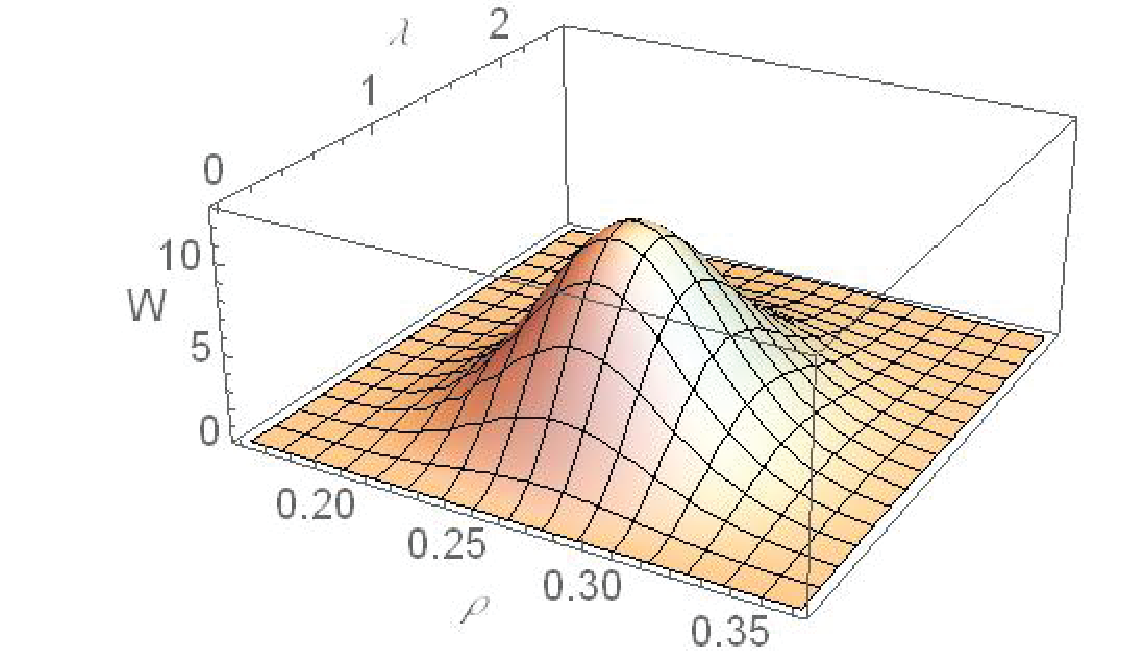}
\includegraphics[scale=0.4]{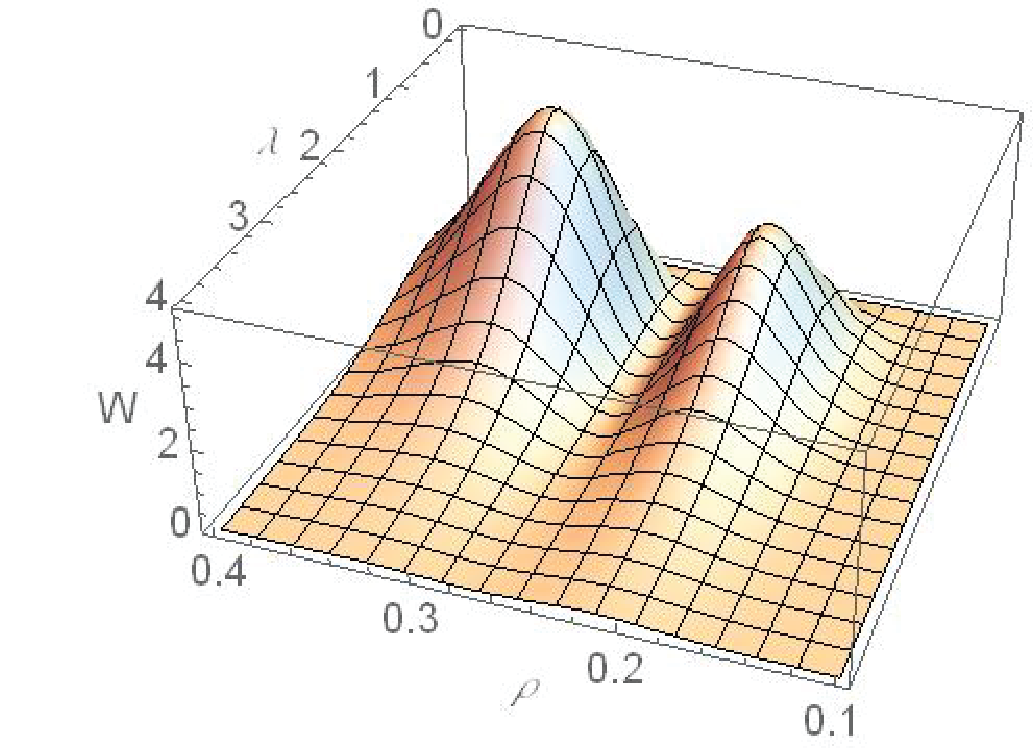}
\caption{The radial distribution density $W(\rho,\lambda)$ for $(K e ^3He)$
in states (21,20) and (21,19).
The variable values $\rho$ and $\lambda$ are taken in electron atomic units.
}
\label{pic6}
\end{figure}

The radial distribution densities are presented in Fig.~\ref{pic1}, \ref{pic2} in the case of pionic helium
and kaonic helium.
These plots show the presence of characteristic distances in the particle systems
$(He-\pi^--e)$ and $(He-K^--e)$.
It also follows from these graphs that for the considered states, the meson turns out 
to be located at the same distances from the nucleus or slightly closer to the nucleus then the electron.
The distribution densities for the two radial variables $\rho$ and $\lambda$ provide a more complete 
picture of the characteristic distances in a given system of three particles. They are shown in two graphs in 
Fig.~\ref{pic5},\ref{pic6}.

Fine splitting in a three-particle atom which is determined by the interaction of the electron spin and the 
large orbital angular momentum of the meson is not considered here.

\section{Discussion of the results}

This paper examines the energy levels of pionic and kaonic helium for states in which the meson has 
such a large orbital momentum that it is located approximately at the same distance from the nucleus as the electron.
The calculations are performed in leading order within the framework of the variational method with the Gaussian basis, 
and a number of basic corrections determined by the Breit Hamiltonian (for relativism, 
vacuum polarization and contact interaction) are calculated in the first order of perturbation theory.
Since an electron is in the $1S$ state, the notation $(n,l)$ is used for a state of three particles 
in Table~\ref{tb1}, where $l$ is the orbital momentum of the meson, and $n$ is the principal quantum number 
for the subsystem $(\pi^- He^{2+})$, $(K^- He^{2+})$.

The Rydberg states in atoms play an important role in refining the values of fundamental constants. 
Thus, based on the spectroscopy of the Rydberg states in a hydrogen atom, the measurement 
of the Rydberg constant has been improved \cite{H1,H2}. In this problem, by working with the Rydberg states, 
it is possible to eliminate contributions to the structure of a nucleus. In the case of mesonic atoms, 
a use of the Rydberg states makes it possible to reduce the influence of strong interaction on the energy spectrum.

Spectroscopy of various exotic molecules can provide new information about a nature of fundamental 
interactions and the values of fundamental parameters of the Standard Model.
Several years ago, the PiHe collaboration at the Paul Scherrer Institute performed laser 
spectroscopy of the infrared transition 
in three-body pion helium atoms \cite{hori1,hori2}. Such atoms were created in a superfluid (He-II) helium target. 
Similar measurements in antiproton helium atoms embedded in liquid helium were carried out by 
the CERN ASACUSA collaboration \cite{asacusa}.
The antiproton-to-electron mass ratio was determined as $m_p/m_e$=1836.1526734(15) \cite{asacusa}.
The mass of $\pi$ meson can be determined by comparing the experimental transition frequencies in pionic
helium with results of the QED calculation \cite{hori1}.
Although the transition frequency $(17,16)\to (17,15)$ has already been measured for pionic helium \cite{hori2}, 
an analysis of experimental data to extract the pion mass is still ongoing.

Our study of the energy levels of both pionic and kaonic helium is carried out on the basis 
of the variational approach, which was developed in the work \cite{varga}. 
In contrast to the work \cite{hori1}, to describe a three-particle system, the Jacobi coordinates $\boldsymbol\rho$,
$\boldsymbol\lambda$
are used, in which the original Hamiltonian has the form \eqref{eq3}. The second difference between our calculation 
and \cite{hori1} is the use of a Gaussian rather than an exponential basis within the variational method. 
In such a basis, all matrix elements of the Hamiltonian are obtained in a closed analytical form. 
Finally, the third difference between our calculations and \cite{hori1} is that in \cite{hori1}, within the framework 
of the variational approach, the method of complex coordinate rotation is used, and we work with 
a real Hamiltonian and solve the eigenvalue problem \eqref{eq5}.
The obtained numerical results for leading order contribution to the energy of a system and corrections 
to it are presented in Table~\ref{tb1}. Comparing these results with calculation in \cite{hori1}, it is necessary to note 
a slight difference in the results, which appears in the third digit after the decimal point (second coulomb
in Table~\ref{tb1}). For the $(17.16)\to (17.15)$ 
transition frequency for pionic helium-4 that has been measured, our result 180772 GHz is slightly different 
(near one per cent) from the result $183681.5\pm 0.5$ GHz obtained in \cite{hori1} and from experimental value, 
which is 183760(6)(6) GHz.
Our result for similar transition frequency in pionic helium-3
is equal 180594 GHz.
We also present here the results for the transition frequencies $(21,20)\to(21,19)$ 
in the case of kaonic helium. They are equal 69094 GHz $(K-e-{^3_2}He)$ and 67017 GHz $(K-e-{^4_2}He)$.
In general, our results using the Gaussian trial functions (third column of Table~\ref{tb1})
are consistent with calculations with an exponential basis in \cite{hori1} (second column of Table~\ref{tb1})
in the case of pionic helium. In the case of kaonic helium the obtained results are new.
The difference in results is due, in our opinion, to differences in the used variational approaches
in this work and in \cite{hori1} and bases for variational wave functions.

A study of characteristic distances at which the nucleus, meson and electron are located relative 
to each other is shown in Figs.~\ref{pic1}-\ref{pic6} for some states for which the binding energies are calculated. 
The meson is in an excited state with a large orbital momentum $l$. The key parameter with which 
you can estimate its distance to the nucleus is determined by the expression $\sqrt{\mu_1/m_3}$,
where $\mu_1$ is the reduced mass of the meson-nucleus system and $m_3$ is the electron mass.
When the principal quantum number $n=\sqrt{\mu_1/m_3}\approx 16$ for the $(\pi^- {_2^4}He)$ or 
$(\pi^- {_2^3}He)$ subsystems, a movement of the 
$\pi^-$- meson occurs at approximately the same distances from the nucleus and with the same binding 
energy as for an electron.
In the case of kaonic helium, the value of the principal quantum number increases due to 
an increase in the meson mass and reaches the value $n\approx 29$.
This parameter determines the order of the principal quantum number $n$, at which the meson and electron 
have close orbits. But in this work we have so far considered slightly smaller values $n\approx 20$, so that 
the $K^-$-meson is located a little closer to the nucleus.
It follows from Figs.~\ref{pic1}-\ref{pic6} that in a case of the considered Rydberg states of the $\pi^-$ ($K^-$)-meson, 
characteristic distances along $\rho$ and $\lambda$ have close values. 
So, for example, the root mean square value of $\sqrt{\lambda^2}$ for the state (17,16) in $(\pi-e-{_2^4}He)$ is 
60050 fm, and the root mean  square value of $\sqrt{\rho^2}$ for the same state is equal 37210 fm.
This means that the use of an analytical 
method for calculating energy levels as in \cite{phys2023,apm2022} is difficult, since the characteristic series for 
the parameter $M_e/M_\mu$ from \cite{phys2023,apm2022} is not rapidly converging.

When calculating relativistic effects, we take into account only corresponding correction 
for the electron, meaning that the electron is lightest particle in this system, 
and with an increase in the principal quantum number $n$, the orbital speed is determined by the formula 
$v=Z\alpha/n$. Therefore, for a meson in the circular Rydberg states it is suppressed by the factor $n$.

In Table~\ref{tb1} we limited ourselves to presenting numerical results of calculating the energies 
of bound states of three particles only for a certain number of states with $(n,l)$. But obtained general 
analytical formulas for the matrix elements of the Hamiltonian of a system make it possible 
to carry out corresponding numerical calculations for other states $(n,l)$, which may be more important 
for the experiment.
For the principal quantum number $n=29$, the binding energy of kaonic helium in the state $(n,l)=(29,28)$ 
is equal to -2.8001942461 e.a.u., and in the state $(n,l)=(29,27)$ it has the value -2.9152046696 e.a.u., which ultimately 
gives the transition frequency between these levels $\nu=756732$ GHz.

\begin{acknowledgements}
This work is supported 
by Russian Science Foundation (grant No. RSF 23-22-00143).
\end{acknowledgements}

\end{document}